 \documentclass[final,5p,times,twocolumn,authoryear]{elsarticle}

\usepackage{amssymb}
\usepackage{amsmath}
\usepackage{orcidlink}

\usepackage{xcolor}

\journal{Physics Letters B}

\begin{document}

\begin{frontmatter}


\ead{ziad.sakr@net.usj.edu.lb}

\title{Applying the BF method on the DESI evidence for dynamical dark energy models }


\author[first,second,third]{Ziad Sakr \orcidlink{0000-0002-4823-3757}  }
\affiliation[first]{Instituto de Física Teórica UAM-CSIC, Campus de Cantoblanco, 28049 Madrid, Spain}
\affiliation[second]{Institute of Theoretical Physics, Philosophenweg 16, Heidelberg University, 69120, Heidelberg, Germany}
\affiliation[third]{organization={Saint Joseph University, Faculty of Sciences},
            city={Beirut},
            postcode={BP-11514}, 
            country={Lebanon}}

\begin{abstract}
Recent baryon acoustic oscillation (BAO) measurements from the Dark Energy Spectroscopic Instrument (DESI), when combined with cosmic microwave background (CMB) data and Type Ia supernovae (SN) observations, indicate a preference for dynamical dark energy, specifically when considering the Chevallier-Polarski-Linder (CPL) model, over the standard $\Lambda$CDM or the $w$CDM model. However, the Bayes factor, a key metric for model comparison, remains inconclusive on which model is preferred. Moreover, its outcome may be affected by the prior ranges that are considered. This paper applies a recently introduced model comparison test, dubbed BF method, that integrates both Bayesian and frequentist approaches to DESI data to address the limitations of purely frequentist methods, such as the lack of prior information or restrictions in the exploration of the full posterior distribution; and purely Bayesian methods, which are sensitive to prior choices and rely on subjective scales to compare models. The method consists in considering the Bayes factor as a random variable and calculates its distribution, that results from values computed in a frequentist approach after perturbing the data following the model considered. It then compares the $p$-values, obtained for each model supposing the null hypothesis is the Bayes factor calculated with the actual data, without the need to consider a qualitative scale.
We apply this hybrid method to DESI data, comparing the CPL and $w$ models under various prior conditions, including weak and strong priors, and theory-informed priors. We find that, when the traditional bayes factor is considered, that weak priors favor the $w$ model over CPL, while strong priors favor CPL. Additionally, theory-informed priors further enhance the preference for the $w$ model. While when we apply the BF method, the preference for CPL over $w$ is seen in all cases albeit with similar but reduced impact on the $p$-value from the different prior considerations. We also tried to generalize the method further, by perturbing as well the covariance matrix following the model considered, and found that, in general, the current data in that case is not stringent enough to disentangle between the two models.
Our results demonstrate that varying the Bayes factor as a random variable, providing that the covariance matrix is kept as model independent, provides a robust model comparison, reducing the impact of prior dependence as well as offering quantitative assessment of the preferences of the competing models.

\end{abstract}



\begin{keyword}
Bayesian \sep frequentist \sep model comparison \sep dynamical dark energy



\end{keyword}

\end{frontmatter}




\section{Introduction}
\label{sec:intro}

Recent results from the Dark Energy Spectroscopic Instrument (DESI) have intensified scrutiny of statistical comparisons between $\Lambda$CDM and dynamical dark‑energy models. Several analyses combining DESI BAO with CMB and supernova data report mild preferences for evolving dark energy.
But because the strength and direction of these preferences depend sensitively on analysis methodology and prior choices, these findings have prompted careful follow-up discussion regarding the robustness of the inferred dynamics and the model-comparison statistic \cite{Adame:2024a,Adame:2024b,Patel:2024,Sakr:2025,Cortes:2024,Liu:2024}. One concern is that combining constraints from probes that occupy partly discrepant regions of parameter space (CMB, SN, and BAO) can produce an apparent preference for dynamics, and that some entries in the DESI24 compilation may be outliers relative to other datasets \cite{Liu:2024,Sapone:2024ltl}. A second, independent issue is that Bayesian model comparison via the Bayes factor (the evidence ratio) is central to these claims. The Bayes factor penalizes model complexity through prior volume, yet it can be sensitive to prior choices; in cosmology—where physically motivated priors for phenomenological parameters (e.g., w0, wa) are limited—this sensitivity can change conclusions qualitatively \cite{Amendola:2024,Nesseris:2013,Efstathiou:2004}. Since the evidence is the likelihood integrated over the prior volume, such that the adopted prior ranges for $w_0$ and $w_a$ can substantially influence the Bayes-factor outcome even when priors are intended to be uninformative. Several recent works have emphasized that priors and parametrization choices (including uninformative versus theory‑informed priors, linear versus physically motivated transforms, and choices of pivot redshift) alter the effective Occam penalty and therefore the evidence ratio between $\Lambda$CDM and dynamical models and show that extending or altering prior ranges for the CPL parameters can change both the sign and magnitude of the evidence preference between CPL and $\Lambda$CDM \cite{Patel:2024,Cortes:2024,Sakr:2025}.

To mitigate, but also beyond, prior sensitivity, recent theoretical progress provides analytic control over how Bayesian evidence itself fluctuates across finite data realizations. In the Gaussian/linear approximation (Gaussian likelihoods, linear signal dependence on parameters), closed‑form expressions and asymptotic formulae for the sampling distribution of the Bayes ratio have been derived \cite{Amendola:2024,Keeley:2021,Good:1957,Leclercq:2023} (see also \cite{Keeley:2025rlg} recent study for a test on dynamical energy evidence but based on the distribution of the goodness of fit quantity). These results link Bayesian evidence to frequentist test statistics by showing that, under well‑specified approximations, the log Bayes ratio follows known distributions whose variance and bias depend on the true parameter offsets and prior widths. Analytic treatments also clarify how evidence for a more complex model can be driven by statistical fluctuations of the data rather than genuine signal, and how Occam factors trade parameter‑fit gains against prior volume penalties in finite samples \cite{Amendola:2024,Cowan:2024}.

The paper is organized as follows. Section 2 summarizes the BF formalism, prior‑volume effects, and analytic distributional results in the Gaussian/linear regime. Section 3 describes the DESI datasets, likelihood approximations, parametrizations, and prior families we consider, together with numerical evidence‑estimation methods. Section 4 presents the results from sampling‑distribution calibrated significances and dependence of evidence on the priors or the perturbation of the data methods, before we conclude in Section 5 with a summary of findings and recommendations for future DESI releases and community standards for model‑comparison reporting.

\section{Theoretical formalism and expectations}

Bayes' factor  is used in general to compare two competing models, where the fiducial model is compared to a new one by computing the ratio $B$  defined as 
\begin{equation}B(A,B)= \frac{E({ D}|{ M}_A)}{E({ D}|{ M}_B)}, 
\end{equation}
where $E$ is the evidence 
\begin{equation} \label{eq:EvidB} E({ D}|{ M}) = \int d \theta\ { L}({ D}| \theta,{ M})\, { P}( \theta,{ M})\,,
\end{equation}
with ${L}(D|M_{A , B})$
the probability of the data $D$ given model $M_{A , B}$ with parameters $\theta_\alpha$ and a prior ${\cal P}$. Model $M_A$ is favored over model $M_B$ if $b >  1$, and $M_B$ is favored otherwise. Bayes' theorem
arises from the fundamental definition of conditional probability
\begin{equation}
P(A|B)=\frac{P(A,B)}{P(B)}
\end{equation}
so that we can write for the joint probability of data and theory
\begin{equation}\label{eq:PLT}
P(D,T)=L(D|T)\pi(T)
\end{equation}
where we see that both data and parameters are considered
random variables. 
This joint probability is the basic object of the FB (Frequentist-Bayesian)
approach. If we marginalize over $T$ in Eq.~(\ref{eq:PLT}) we get the evidence
\begin{equation}\label{eq:ELT}
\int L(D|T)\pi(T)dT=E(D)
\end{equation}
where the data are still random variables. So we can see $E(D)$ both
as a frequentist in Eq.~(\ref{eq:ELT}) and as a Bayesian object as in Eq.~(\ref{eq:EvidB}) In the former case, we
can vary $D$ and find the distribution of $E(D)$ or of the ratio
$R\equiv E(D|A)/E(D|B)$; in the Bayesian case, the ratio $R$ is
a fixed number and directly expresses the odds of A versus B.

To gain insights we consider the instructive case of Gaussian data with linear nested models,
Bayes' ratio for models $A$ and $B$ (with $N_{A}$ and $N_{B}$ free parameters), $F$ the fisher matrix when considering either model, and the same applies for $L$ the likelihood, $P$ the prior probability, $\hat{\theta}$ vector of best-fit parameters, $\tilde{\theta}$ vector of prior means and $\chi^2$ goodness of fit . $B$ is then 
\begin{align}
B & =\frac{E_{A}}{E_{B}}=\frac{|\Sigma|^{-\frac{1}{2}}e^{-\frac{1}{2}{\hat\chi}_{A}^{2}}\frac{|P_{A}|^{1/2}}{|F|_{A}^{1/2}}\exp\left[-\frac{1}{2}(\hat{\theta}_{A}-\tilde{\theta}_{A})L_{A}F_{A}^{-1}P_{A}(\hat{\theta}_{A}-\tilde{\theta}_{A})\right]}{|\Sigma|^{-\frac{1}{2}}e^{-\frac{1}{2}\hat{\chi_{B}}^{2}}\frac{|P_{B}|^{1/2}}{|F_{B}|^{1/2}}\exp\left[-\frac{1}{2}(\hat{\theta}_{B}-\tilde{\theta_{B}})L_{B}F_{B}^{-1}P_{B}(\hat{\theta}_{B}-\tilde{\theta}_{B})\right]}\\
 & =e^{-\frac{1}{2}(\hat{\chi}^{2}_{A}-\hat{\chi}_{B}^{2})}\frac{|P_{A}F_{B}|^{1/2}}{|P_{B}F_{A}|^{1/2}}\exp\left[-\frac{1}{2}(\chi_{P,A}^{2}-\chi_{P,B}^{2})\right]
\end{align}
or 
\begin{align}\label{eq:fullbeta}
\beta & \equiv2\ln B=\chi_{B}^{2}-\chi_{A}^{2}+\chi_{P,B}^{2}-\chi_{P,A}^{2}+Y_{B}-Y_{A}
\end{align}
 where
\begin{equation}
Y_{M}=\ln\frac{|F_{M}|}{|P_{M}|}
\end{equation}
and
\begin{equation}
\chi_{P,M}^{2}=(\hat{\theta}_{M}-\tilde{\theta}_{M})L_{M}F_{M}^{-1}P_{M}(\hat{\theta}_{M}-\tilde{\theta}_{M})
\end{equation}
with $M=A,B$.
In Eq.~(\ref{eq:fullbeta}), the last term, expresses the deviation between prior and posterior
peak (Occam penalty), defined as the gain, or improvement on our knowledge (volume of constraints)
when going from the prior to the posterior,
 while the first term gives the distance between the peaks, and the second how close are the best-fits to the original priors. 

In the standard Bayesian approach, model A is preferred if $\beta$, calculated once with fixed data value, is positive and large, that is, when it is a good fit (small $\chi^2_A$), which does not depart too much from the prior (small $\chi^2_{P,A}$), and when the gain from the prior to the final uncertainty is minimal  (small $Y_A$).
However, especially for nested models, as is the case here, and as we shall see in the result section later,  \cite{Amendola:2024} has showed that the model with higher number of parameters is more disfavoured when wider priors are considered, which makes a comparison based on the fixed $B$  strongly prior dependent.

While here in the BF method, $\beta$ would vary since we vary the data, and instead of competing $\chi^2$ and $Y$ terms for fixed data, we compare the odds from the distribution we get when we elevate $B$ to a random variable and vary the data to obtain two distribution assuming either model A or model B is true. From Eq.~(\ref{eq:fullbeta}) we might then expect to see that a `robust' rejection of one model (i.e. a rejection that holds up under $n$ repeated $i$ draws of the data) depends on the relative magnitude of the $(\chi^2_M)_i$ terms with respect to the $(Y_M)_i$ terms since each model would minimize its  $(\chi^2_M)_i$ terms, but if $Y_{M,i}$, which depends on the experiments measurement precision with respect to the its prior range, is $ \gg \chi^2_{M,i}$, then $\beta$ will hardly scatter under repetitions of the experiment. A good indication that is not sufficient however to conclude on the comparison of both since, apart from the scatter relative to each model, we are also interested in the final $p$-value of each distribution. The latter in the BF method would be calculated choosing the fixed $B$ (assuming no model is true) as our limit to the integral giving $p$. We are now by then comparing the differences between the likeliness of each distribution with respect to the null hypothesis. For linear priors, this difference will essentially stay the same such that the dependence on it end up not affecting our comparison and we have by then limited the prior influence, which will allow us to better perform our model comparison.

We finally note that in our method above, only the data measurements are perturbed following the model we consider each time to construct the Bayes' factor distribution. However, the method could be further generalised, as we shall see later, to also perturb the covariance matrix as well. While in the first procedure, we expect that each distribution will naturally peak around a $B$ value that favours the model we have chosen to perturb the data, with a larger variance for the model with more degrees of freedom (dof) (we label as model B) which is even more expected in our nested models case, since it should at least still allow values of $B$ compatible with those of the small dof model (here model A); this would not be exactly the case in the further generalisation, since perturbing the errors themselves would weaken the constraints of the data resulting in: 
a reduction of the distance between the peaks of the $B$ distribution since now each model will less sharply best fit its perturbed data more than the other; and a comparable value of the variance when considering either of model A or B, since, as explained previously and without lose of generality when using the simple example to gain insight, a scatter comes from an equal interplay between $Y_M$ and $\chi^2_M$ for each model, therefore, for model A an increase of $F_M$ will make it now less dominated by $\chi^2_A$, while for model B, $Y_B$ increase will make it dominate $\chi^2_B$ thus reducing its scatter. 

\section{Implementation methods and datasets}
\label{sec:meth&data}

First, we run MCMC chains to obtain the best-fit and variance on the cosmological parameters for either the $w$ or CPL ($w_0w_a$) model using the real DESI, Pantheon + data and compressed CMB background evolution $\theta^*$ (angular size of the sound horizon at the time of recombination) prior. We use these values as the average and variance of a Gaussian distribution to draw from it different sets of the parameter values. For each new set, we rescale the data with the new theoretical predictions for our observables evaluated at the redshifts of the actual data points and run MCMC chains again in order to be used in the calculation of the different values to generate an ensemble of Bayes factor evaluations. Since the subsequent MCMC runs are so computation and time costly, which force us to limit our sample to realistic size, we adopt a Latin hypercube method to generate our sample of parameter values from a multidimensional Gaussian distribution centred around our best fit parameters and their variances in order to optimize the coverage of the space along with limiting overlapping redundant areas. Practically, the procedure is repeated multiple times (typically $O\,(10^3 )$). As mentioned, to obtain each value of our sample, we perform an MCMC run for each of our cosmological and nuisance (for SN) parameter sets for both the CPL ($w_0w_a$) and $w$ models. We then use the \texttt{MCEvidence} code to compute the needed evidence $E$ to calculate $B$ and plot the resulting Bayes factor distribution when varying the data considering each of the models along with the line indicating the value of $B$ for the original data. We calculate and plot $B$ distributions considering in first case the same weak priors for both models $w_0w_a$ and $w$ in comparison to adopting same strong priors for both. We then consider different weak priors for each model stemming from theoretical considerations such as $w$ cannot go above $-1/3$ to still allow an accelerating universe. In a third and forth case, we repeat the priors scheme used above but we now perturb the covariance matrix of the data as well. 

\begin{figure}
	\centering 
	\includegraphics[width=\columnwidth, ]{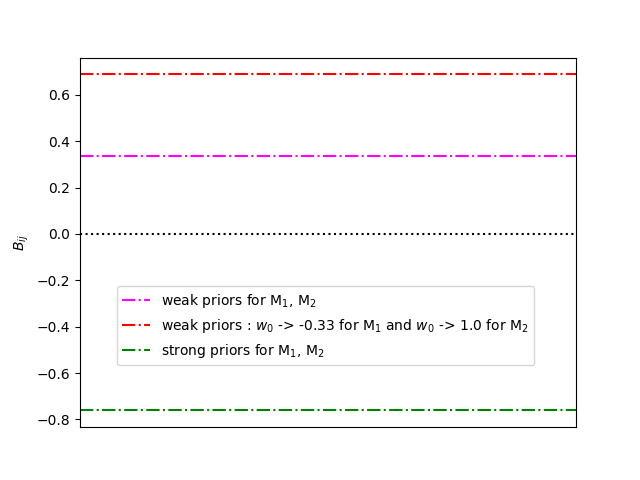}	
	\caption{Bayes factor for the $w$ and $w_0w_a$ models evaluated when considering different priors using Planck 2018 CMB geometrical compressed information, Pantheon+ Supernova sample and DESI BAO measurements.} 
	\label{fig:Bfix}%
\end{figure}

\begin{figure}
	\centering 
	\includegraphics[width=\columnwidth, ]{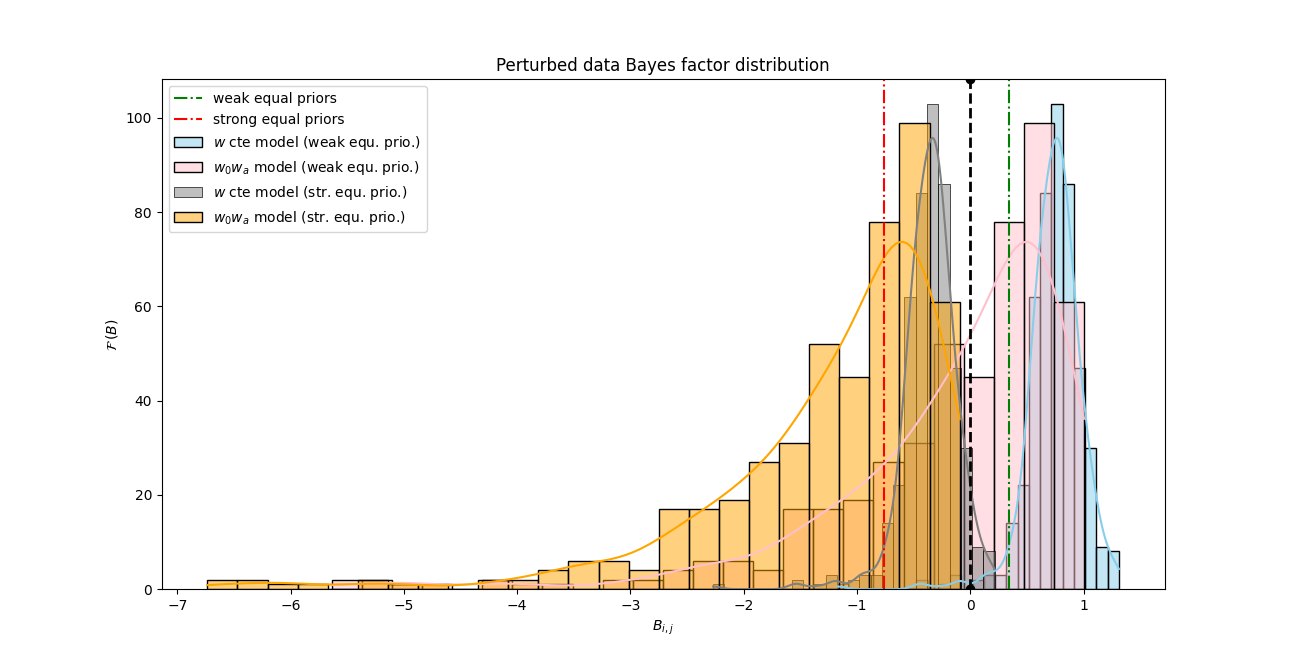}	
	\caption{Distribution of $B$ histogram bars and kernel smoothed interpolation solid line assuming either model $w_0w_a$ or model $w$ with equal strong or weak prior for both models (see the legend for details). The vertical lines corresponds to the value of $B$ when the data is not perturbed. Settings are the same as Fig.~\ref{fig:Bfix}} 
	\label{fig:Bvarprioequ}%
\end{figure}

\begin{figure}
	\centering 
	\includegraphics[width=\columnwidth, ]{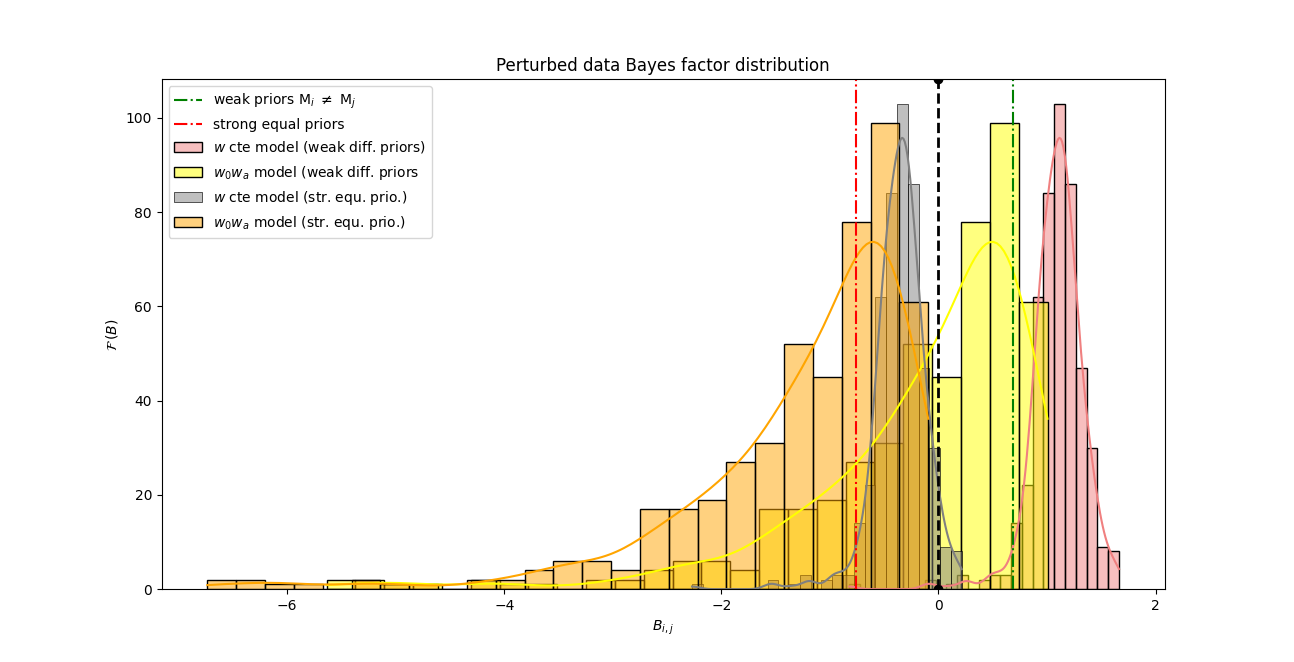}	
	\caption{Distribution of $B$ histogram bars and kernel smoothed interpolation solid line assuming either model $w_0w_a$ or model $w$ with equal strong for both models but different theory inspired weak priors (see the legend for details). The vertical lines corresponds to the value of $B$ when the data is not perturbed. Settings are the same as Fig.~\ref{fig:Bfix}} 
	\label{fig:Bvarpriodiff}%
\end{figure}

\begin{figure}
	\centering 
	\includegraphics[width=\columnwidth, ]{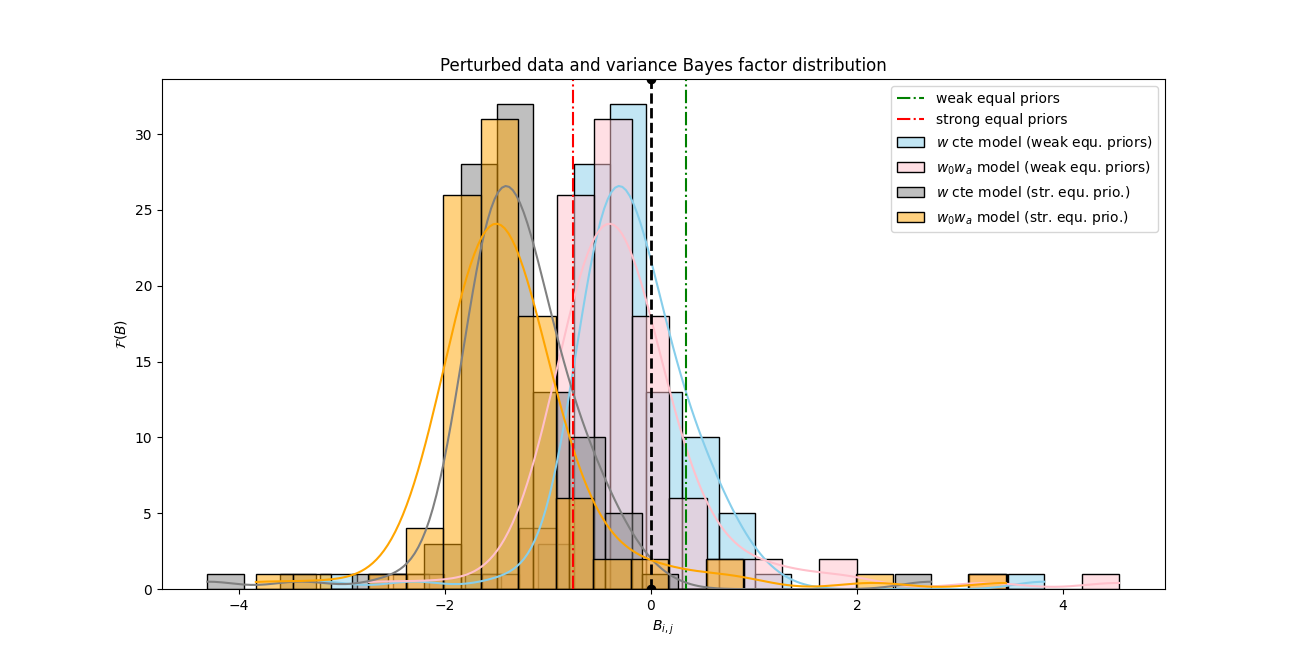}	
	\caption{Distribution of $B$ histogram bars and kernel smoothed interpolation solid line assuming either model $w_0w_a$ or model $w$ with equal strong or weak prior for both models (see the legend for details) perturbing the data and the their covariance matrices. The vertical lines corresponds to the value of $B$ with fixed data. Settings are the same as Fig.~\ref{fig:Bfix}} 
	\label{fig:Bvarcovprioequ}%
\end{figure}

\begin{figure}
	\centering 
	\includegraphics[width=\columnwidth, ]{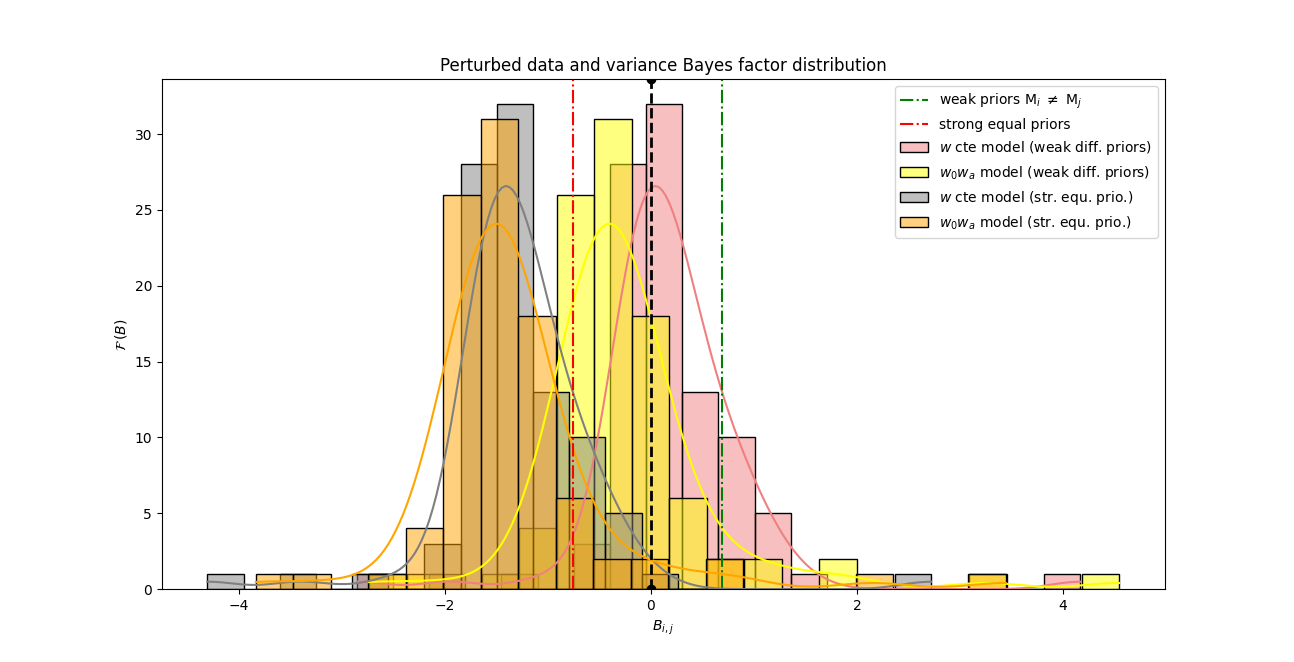}	
	\caption{Distribution of $B$ histogram bars and kernel smoothed interpolation solid line assuming either model $w_0w_a$ or model $w$ with equal strong for both models but different theory inspired weak priors (see the legend for details) perturbing the data and the their covariance matrices. The vertical lines corresponds to the value of $B$ with fixed data. Settings are the same as Fig.~\ref{fig:Bfix}} 
	\label{fig:Bvarcovpriodiff}%
\end{figure}

\section{Results}
\label{sec:res}
We start by looking at the classical fixed $B$ in Fig.~\ref{fig:Bfix} where we observe as expected that weak priors favor $w$ over $w_0w_a$ while adopting strong priors favours $w_0w_a$. We see then the effects when using weak priors. Moreover, we are considering the same priors for both models. If we use theory informed prior for w model, since the latter value cannot be $<$ 1/3, then we see as expected, that w is even more favoured since we reduced the diluting volume. 

Then we see how varying $B$ as in Fig.~\ref{fig:Bvarprioequ} changes our conclusion and favours $w_0w_a$, first in the case of strong prior, a case that coincide with the fix $B$ one, but most importantly in the case of weak prior we notice almost a non change in the preference for our competing models, between strong and weak priors, while we see in Fig.~\ref{fig:Bvarpriodiff}, where we consider different weak priors, the $w_0w_a$ becomes less favoured, albeit still beating $w$. That translates into the $p$-value in table \ref{tab:pvalue} where in both scenarios the $w_0w_a$ corresponding value remains higher than $w$ model, but it decreases in the weak different prior cases, where the difference is not that big between $w_0w_a$ and $w$ model. 

In the case of full propagation of errors, it is expected that the scatter in the data should reduce the power of the constraints and thus favour the model with less degrees of freedom since it can now equally fit the data close to the power of the more sophisticated model. Thus we observe in Fig.~\ref{fig:Bvarcovprioequ} that the line of fixed B lies now, whether in the weak or strong same priors for both models, within $2~\sigma$ in for each model, which means that the $w$ model has become less rejected while model $w_0w_a$ has become less favoured. While in the case of different equal priors in the weak limit we show in Fig.~\ref{fig:Bvarcovpriodiff}, both effects, scatter in the data and volume dilution, would combine to make the $w_0w_a$ case much less favoured than the $w$ model. This translates in the a drop of the $p$-value in table \ref{tab:pvalue} of $w_0w_a$ below that of $w$ making the later now favoured in the weak limit even in the varied B method, while in the case of the equal priors, the $p$-value of $w_0w_a$ is now close to that of $w$

\begin{table}\label{tab:pvalue}
\begin{center}
\setlength{\tabcolsep}{5pt}
\renewcommand{\arraystretch}{1.55}
\caption{$p$-values for the dark energy extension models. }
\begin{tabular}{ccc}
\hline
\hline
Models                       & $w_0$   & $w_0w_a$ \\ \hline
equal weak or strong priors            & $0.04$ & $0.60$ \\ \hline
weak different priors             & $0.04$ & $0.19$ \\ \hline
equal weak or strong priors error propagation            & $0.17$ & $0.15$ \\ \hline
weak different priors error propagation           & $0.17$ & $0.09$ \\ \hline
\end{tabular}
\end{center}
\end{table}

\section{Summary and conclusions}
\label{sec:conc}
We investigated the robustness of Bayesian model comparison between $w$ model and dynamical dark‑energy CPL ($w_0w_a$) parametrization in light of recent DESI BAO results combined with CMB and supernova data. Motivated by prior-sensitivity concerns and by recent analytic work on the sampling distribution of the Bayes factor $B$ in the Gaussian/linear regime, we performed an end-to-end numerical experiment that estimates the sampling distribution of $B$ under repeated data realizations.

For that, we generated synthetic data realizations by drawing cosmological parameters from Gaussians centred on MCMC-derived best fits from BAO from DESI, SN from Pantheon+ sample, and compressed CMB priors, then rescaling observables befotre re-running MCMC to compute a set of evidences with the \texttt{MCEvidence} code. We explored two prior schemes: one considering same weak or strong priors for both models; and another with different theory‑informed (for the single-parameter $w$ model) weak priors. For each scheme we considered two data-perturbation approaches: perturbing only data realizations while holding the covariance fixed, and another where we also perturb the covariance (full propagation of errors).

We find that fixed $B$ (single evidence) conclusions are highly prior dependent: weak, broad priors tend to prefer the simpler $w$ model over $w_0w_a$, whereas tighter (strong) priors can reverse preference toward $w_0w_a$ because of reduced Occam penalty.  While when the Bayes factor is treated as a random variable (sampling distribution approach), some prior influence persisted but conclusions change. In the case of equal strong priors for both models, sampling fluctuation results agree with the fixed $B$ on preferring $w_0wa$ over $w$. The conclusions stays the same for weak priors unlike it was the case with the fixed $B$ method. Using different weak priors in the BF method reduces CPL’s advantage and declines the $p$-value though CPL still attains a higher median Bayes factor than $w$. We also find that propagating covariance perturbations reduces the method discriminating power with current data, with the scatter in the data and in the covariance now further weakening constraints, resulting in an overlap region of the sampling distributions of the weak or strong cases where $w_0w_a$ is not strongly favoured as before, to the extent that, in the asymmetric prior weak case, the simpler $w$ model becomes equally favoured based on the $p$ values.

We conclude that current mild evidence for evolving dark energy from DESI+external datasets, not yet decisive with the usual Bayes factor method, is however also showing similar preference for $w_0w_a$ over $w$ (the models we studied here) in almost all the variants considered with the more robust Bayesian Frequentist method. We however recommend applying the BF method to the upcoming DESI or other Stage-IV releases in order to improve robustness and interpretability of Bayes factor claims, and reduce spurious inferences of dynamics and better identify genuine deviations from $\Lambda$CDM. This is even more recommended and justified from the most general case we considered, where we perturbed the covariance matrix as well, where an increase of the quality and precision of the data, will remedy for the reduction of the difference of the $p$ value and establish back a clear preference for one of the models over the other.

\section*{Acknowledgements}
ZS acknowledges support from DFG project 456622116 at the time when this study was conducted and support from the research projects PID2021-123012NB-C43, PID2024-159420NB-C43, the Proyecto de Investigación SAFE25003 from the Consejo Superior de Investigaciones Científicas (CSIC), and the Spanish Research Agency (Agencia Estatal de Investigaci\'on) through the Grant IFT Centro de Excelencia Severo Ochoa No CEX2020-001007-S, funded by MCIN/AEI/10.13039/501100011033.

\bibliographystyle{elsarticle-num-names} 
\bibliography{example}

\end{document}